\documentclass[12pt,a4paper,english,fleqn]{article}

\usepackage[sort&compress,round,comma,authoryear]{natbib}
\usepackage[T1]{fontenc}
\usepackage{ae,aecompl}

\usepackage[colorlinks=true,urlcolor=blue,citecolor=blue,linkcolor=red,bookmarks=true]{hyperref}
\usepackage{graphicx}	
\usepackage{float}	% Including figure files
\usepackage{amsmath}	% Advanced maths commands
\usepackage{amssymb}	% Extra maths symbols
\newlength{\lp}
\usepackage[mathscr]{eucal}
\usepackage{amsfonts}
\makeatother
\usepackage{babel}
\makeatother
\begin{document}
\title{MOND: A consequence of the geometric Leibniz Clock}
\author{D. F. Roscoe (The Open University; D.Roscoe@open.ac.uk)\\ \\ORCID: 0000-0003-3561-7425}
\date{}
\maketitle
\newpage
% Abstract of the paper
\begin{abstract}
Leibniz considered the notion of the \emph{empty physical space} to be a meaningless abstraction, and he held firmly to the view that the only significant thing was the set of relationships between \lq{objects}', whatever these \lq{objects}' might be. 
Similarly, he was equally clear in expressing his views about Newton's \emph{universal time}, which he also considered to be a meaningless abstraction. In effect, for him, \emph{time} was no more than a synonym for ordered change within a material system. 
\\\\
The process of giving quantitative realization to this duality of non-Newtonian ideas forms the core of this work. A primary result arising is that every gravitating particle is no more than a \emph{clock} - the geometric Leibniz Clock - which provides all the basic things: it conserves energy and angular momentum and satisfies the Weak Equivalence Principle. 
\\\\
When the Clock is applied to model the concept of a galactic object within which all motions are circular, the characteristic properties of the MOND galaxy (asymptotic flatness, a critical acceleration scale, the baryonic Tully-Fisher relationship) are quantitatively reproduced in the resulting Leibniz galaxy. In short, the characteristic essence of MOND has its source in the geometric Leibniz Clock.  
\end{abstract}
\newpage{}
% Select between one and six entries from the list of approved keywords.
% Don't make up new ones.
%\begin{keywords}
%MOND -- Fractal -- homogeneity
%\end{keywords}

%%%%%%%%%%%%%%%%%%%%%%%%%%%%%%%%%%%%%%%%%%%%%%%%%%

%%%%%%%%%%%%%%%%% BODY OF PAPER %%%%%%%%%%%%%%%%%%
\section{Introduction: }
The correspondence of Clarke-Leibniz (1715$\sim$1716) (\citet{Alexander1984}) concerning the nature of physical space and experienced time  makes it clear that Leibniz considered the concepts of the \emph{empty physical space} and (Newton's) \emph{universal time} to be meaningless abstractions. In modern terms, for Leibniz,  \emph{physical space} is a secondary construct projected out of primary relationships between \emph{objects} (whatever these might be) whilst, for him, \emph{time} stands as no more than a synonym for \emph{ordered change} within material systems.
\\\\
These ideas are simple to state, and might even seem self-evident to some, but they are very difficult to realize in any quantitative way. To see why this might be, it is instructive to consider the case of \emph{physical space}:  if this is a secondary construct projected out of primary relationships between objects then, of itself, it stands as a proxy for its own contents - they are one and the same thing. This is a deeply recursive relationship, and therein lies the difficulty associated with any attempt to synthesize a quantitative Leibnizian worldview.
\\\\
But therein also lies the solution: once the necessarily recursive nature of a realized Leibnizian worldview is explicitly recognized, then the primative key required for its synthesis is readily found. But once synthesized, this recursive nature means that its quantitative features can only be easily discussed through the prism of a classical interpretation: that is, by using terms such as \emph{geometry, metric structure, spatial distribution, dynamics, equilibrium, time} and similar. As a primary result, it is found that every gravitating particle is no more than a clock - the geometric Leibniz Clock - which provides all the basic things: it conserves energy and angular momentum and satisfies the Weak Equivalence Principle.
\\\\
But beyond these basic things, there is much more: in particular, there is a basic equilibrium state in which global material is necessarily distributed in a $D=2$ fractal fashion and such a distribution is, by definition, characterized by a mass surface density scale $\Sigma_F$ or, equivalently, by a characteristic acceleration scale $a_F \equiv 4\pi G\,\Sigma_F$.  The existence of such an irreducible equilibrium state is consistent with the claims of \citet{Baryshev1995} and many others over the years, that galaxies on medium scales $(100\sim 300\,Mpc)$ are distributed in a $D\approx 2$ quasi-fractal fashion. See appendix \S\ref{FractalUniverse} for a detailed review of the literature. 
\\\\
On the basis of this evidence, we make the hypothesis that the $D=2$ fractal equilibrium state extends to the finer scales of the intergalactic medium (IGM) itself, and that an isolated galactic object can be modelled as a finite bounded perturbation of this idealized IGM. When motions within the model galaxy are constrained to be purely circular, the irreducible properties of the geometric Leibniz Clock for the system encapsulate exactly the characteristic features attached to the isolated galaxy by Milgrom's MOND (\citet{Milgrom1983a, Milgrom1983b,Milgrom1983c, Milgrom1983d, Milgrom1983e}) - with the detailed difference that, wherever it occurs, MOND's critical acceleration scale $a_0$ is replaced by $a_F$. 
\\\\
A direct calculation using the SPARC data of \citet{Lelli2016A} and the photometric mass-models of \citet{Lelli2016B}, together with the invariance properties of the geometric Leibniz Clock, then yields   
\[
a_F \equiv 4 \pi G \Sigma_F \approx 1.2\times 10^{-10}\,mtrs/sec^2
\] 
showing that Milgrom's critical acceleration scale is explicitly and quantitatively identified as the characteristic acceleration scale of the fractal equilibrium state.  
\\\\
In summary, the essential core of Milgrom's MOND is encapsulated exactly in the basic properties of the most simple possible galaxy model in the geometric Leibniz worldview.
\section{A brief history of ideas of space and time}\label{sub:A-brief-history}
\subsection{Space}
The conception of space as the container of material objects is generally
considered to have originated with Democritus and, for him, it provided
the stage upon which material things play out their existence - \emph{emptiness}
exists and is that which is devoid of the attribute of \emph{extendedness}
(although, interestingly, this latter conception seems to contain
elements of the opposite view upon which we shall comment later).
For \citet{Newton1999}, an extension of the Democritian conception
was basic to his mechanics and, for him:
\begin{quote}
	\emph{... absolute space, by its own nature and irrespective of anything
		external, always remains immovable and similar to itself.}  
\end{quote}
Thus, the absolute space of Newton was, like that of Democritus, the
stage upon which material things play out their existence - it had
an \emph{objective existence} for Newton and was primary to the order
of things. In a similar way, time - \emph{universal time,} an absolute
time which is the same everywhere - was also considered to possess
an objective existence, independently of space and independently of
all the things contained within space. The fusion of these two conceptions
provided Newton with the reference system - \emph{spatial coordinates}
defined at a \emph{particular time} - by means of which, as Newton
saw it, all motions could be quantified in a way which was completely
independent of the objects concerned. It is in this latter sense that
the Newtonian conception seems to depart fundamentally from that of
Democritus - if \emph{emptiness} exists and is devoid of the attribute
of \emph{extendedness} then, in modern terms, the \emph{emptiness}
of Democritus can have no \emph{metric} associated with it. But it
is precisely Newton's belief in \emph{absolute space \& time} (with
the implied virtual clocks and rods) that makes the Newtonian conception
a direct antecedent of Minkowski spacetime - that is, of an empty
space and time within which it is possible to have an internally consistent
discussion of the notion of \emph{metric.}\\
\\
The contrary view is generally considered to have originated with Aristotle  (\citet{Wicksteed} and \citet{McKeon1941})
 for whom there was no such thing as
a \emph{void} - there was only the \emph{plenum} within which the
concept of the \emph{empty place} was meaningless and, in this, Aristotle
and Leibniz (\citet{Ariew}) were at one. It fell to Leibniz, however,
to take a crucial step beyond the Aristotelian conception: in the
debate of Clarke-Leibniz (1715$\sim$1716) (\citet{Alexander1984}) in which
Clarke argued for Newton's conception, Leibniz made three arguments
of which the second was:
\begin{quote}
	\emph{Motion and position are real and detectable only in relation
		to other objects ... therefore empty space, a void, and so space itself
		is an unnecessary hypothesis.} 
\end{quote}
That is, Leibniz introduced a \emph{relational} concept into the Aristotelian
worldview - what we call \emph{space} is a projection of \emph{relationships}
between material bodies (whatever these might be) into the perceived world whilst what we call
\emph{time}, which is implied by the idea of \emph{motion}, is the projection of ordered \emph{change} into the perceived
world. Of Leibniz's three arguments, this latter was the only one to which
Clarke had a good objection - essentially that \emph{accelerated motion,}
unlike uniform motion, can be perceived \emph{without} reference to
external bodies and is therefore, he argued, necessarily perceived
with respect to the \emph{absolute space} of Newton. It is of interest
to note, however, that in rebutting this particular argument of Leibniz,
Clarke, in the last letter of the correspondence (\citet{Alexander1984}), put his finger directly
upon one of the crucial consequences of a relational theory which
Leibniz had apparently not realized (but which Mach much later would)
stating as absurd that:
\begin{quote}
	... \emph{the parts of a circulating body (suppose the sun) would
		lose the} vis centrifuga \emph{arising from their circular motion
		if all the extrinsic matter around them were annihilated.} 
\end{quote}
This letter was sent on October 29th 1716 and Leibniz died on November
14th 1716 so that we were never to know what Leibniz's response might
have been. 
\\
\\
Notwithstanding Leibniz's arguments against the Newtonian conception,
nor Berkeley's contemporary criticisms (\citet{Berkeley1992}), which were
very similar to those of Leibniz and are the direct antecedents of
Mach's, the practical success of the Newtonian prescription subdued
any serious interest in the matter for the next 150 years or so until
Mach himself picked up the torch. In effect, he answered Clarke's
response to Leibniz's second argument by suggesting that the \emph{inertia}
of bodies is somehow induced within them by the large-scale distribution
of material in the universe:
\begin{quote}
	\emph{... I have remained to the present day the only one who insists
		upon referring the law of inertia to the earth and, in the case of
		motions of great spatial and temporal extent, to the fixed stars ...}
	\citet{Mach1960}
\end{quote}
thereby generalizing Leibniz's conception of a relational universe. 
\subsection{Time for Leibniz \& Mach} \label{LeibnizTime}
Leibniz was equally clear in expressing his views about the nature of
time which are very similar to those expressed by Mach.  They each viewed \emph{time} (specifically Newton's \emph{universal time}) as a meaningless abstraction. So, for example, in his exchanges with Clarke (\citet{Alexander1984}) Leibniz famously said
\begin{quote}
 \emph{As for my own opinion, I have said more than once, that I hold space to be something merely relative ... I hold it to be an order of coexistences, as time is an order of successions ...}
\end{quote}
In other words, for Leibniz \emph{time} is no more than a synonym for \emph{ordered change} within a material system.
\\\\
\citet{Mach1960} went further: all that we can ever do, he argued, is to measure \emph{change} within one system against
\emph{change} in a second system which has been defined as the standard
(eg it takes half of one complete rotation of the earth about its
own axis to walk thirty miles).
So, on this Machian view, the `clock' used to quantify the passage of time for a physical system A is simply an independent physical system B which has been arbitrarily defined as the standard clock.
\\\\
It follows that any definition of \emph{time} based upon the Machian argument must be based upon the ideas of: 
\begin{itemize}
	\item every physical system having its own subjective and private internal time-keeping, any one of which can be chosen as the \emph{standard clock} against which all the other systems reckon the passage of time;
	\item in practice, one such system being chosen as the standard clock, the choice being merely one of convention usually, but not necessarily, involving a natural cyclic process.
\end{itemize}
\section{Recursive Leibnizian space}\label{Leibniz}
As we have noted, the debate of Clarke-Leibniz (1715$\sim$1716) (\citet{Alexander1984}) concerning the nature of physical space makes it clear that Leibniz considered the concept of the empty physical space to be a meaningless abstraction, and he held firmly to the view that the only significant thing was the set of relationships between \lq{objects}', whatever these \lq{objects}' might be.
\\\\
As a first step towards quantifying this idea in modern terms, we interpret it to imply the view that 
 \emph{without material content no concept of a metrical physical space can exist}.
\\\\
We then formulate the question: 
\emph{how can one impose metric structure upon a physical space which is such that the metric structure becomes undefined when that physical space is empty?} 
\subsection{Primative key: a  mass-calibrated distance metric} \label{PK}
With the foregoing question in mind, we begin with the universe of our experience, and  make the simplifying assumption that there is at least one point about which the distribution of universal matter is (statistically) isotropic, and that the arguments to follow are based on such a point being at the centre of \emph{an astrophysical sphere}. 
\\\\
The primary step taken is then the recognition that the bigger an astrophysical sphere is, then the more mass it contains, and vice-versa. It follows that an unconventional measure for the radius of such a sphere can be \emph{defined} purely in terms of its mass content, irrespective of the nature of this mass, visible or otherwise. In other words, for such a sphere with a mass content $m$, we can say that its radius is defined by
\begin{equation}
R \underset{\mathrm{def}}{=} G(m) \label{eqn0}
\end{equation}
where $G$ is a monotonic
increasing function of $m$ satisfying only the condition $G(0)=0$ so that the empty sphere necessarily has zero radius. In effect, the idea of a physical space existing independently of its mass content is at once obliterated and so this simple model encapsulates the recursive nature of the relationship between matter and physical space. Note that $R$ only becomes calibrated when the function $G$ is defined.
\\\\
 It follows immediately that an invariant measure of an arbitrary \emph{radial displacement} can be written, purely in terms of mass, as
 \begin{equation}
 \Delta R = G(m+\Delta m) - G(m) \label{eqn1}
 \end{equation}
so that, whatever the form of $G$ chosen, we have a spatial metric
which follows Leibniz in the required sense (that `space' is a secondary construct projected out of a `matter distribution') for any displacement which
is \emph{purely radial.} 
\\\\
The question now becomes: 
\emph{how can we generalize this idea to impose a general metric structure upon a physical space which is such that the metric structure becomes undefined when that physical space is empty?} 
\subsection{Qualitative assessments of `distance' in everyday life}\label{Life}
In order to provide a quantitative answer to this latter question, it is instructive to reflect very briefly upon how we, as primitive human beings, form qualitative assessments of `distance' in our everyday lives without recourse to formal instruments.
\\\\
So, for example, when walking across a tree-dotted
landscape the changing \emph{angular} relationships between ourselves
and the trees provides the information required to assess both
\emph{what distance travelled?} and \emph{which tree nearer/further?} measured in units of human-to-tree angular displacements within that landscape. If we obliterate our view of the scene - say, with fog - then all forms of \lq{distance}' information are destroyed.
\\\\
In other words, the informal metric structure that we impose upon the space containing the landscape derives exclusively from changes in the angular relationships between ourselves and the material elements of that landscape as we move within it. 
This indicates a geometric approach by which a formal metric structure can be imposed upon a specifically non-empty physical space.
\subsection{The mass model, and its generalization from the sphere}\label{MassModel}
Before we can usefully apply the insights of \S\ref{Life}, we need to define exactly what object is to be our \lq tree dotted landscape': 
to this end, we consider the recursive model $R \underset{\mathrm{def}}{=} G(m)$, to be absolutely primary and then invert it to give
the derived mass model, 
\begin{equation}
Mass\equiv m=M(R), \label{eqn0D}
\end{equation}
for any astrophysical sphere in our rudimentary universe. It is this object which we take to represent our \lq tree dotted landscape'.  Note that:
\begin{itemize}
	\item by (\ref{eqn0}), $R$ vanishes in the case of the astrophysical sphere being empty;
	\item $R$ is necessarily non-zero in the case of the astrophysical sphere containing any matter whatsoever, so that the idea of the point-mass source does not exist;
	\item $R$ only becomes calibrated when $G$ becomes defined.
\end{itemize}
In explicit recognition that a 3-space continuum is being admitted, we write (\ref{eqn0D}) as
\[
m \equiv M(R),~~ R \equiv f(x^1,x^2,x^3)
\]
where  $R$ is (as yet) uncalibrated, and we assume nothing about the spatial coordinates, $(x^{1},x^{2},x^{3})$.
\\
\\
Note that, although up to this point, $R=constant$ has been interpreted as a spherical surface for ease of discussion, it can, in fact, represent any topological isomorphism of a spherical surface; for example, an ellipsoid to model an elliptical galaxy; or a thick pancake taken as a primitive model for a spiral galaxy etc. In this broader context, equation (\ref{eqn1}) (originally interpreted to define the idea of a mass-calibrated \emph{radial} displacement) is generalized so that it  defines a mass-calibrated displacement \emph{normal} to the level surface $R=constant$.
\subsection{A mass-calibrated metric for arbitrary spatial displacements}\label{M-definition}
We have a way of assigning a mass-calibrated metric for displacements which are purely normal to the level surface $R=constant$ (of which the sphere is a special case) in our model universe. We now need a way  of assigning a mass-calibrated metric for arbitrary displacements within that universe. The reflections of \S\ref{Life} on how we, as primitive observers in a landscape, manage this without recourse to formal instruments inform our approach to the problem. 
\\
\\
Since we are taking the mass-model $m \equiv M(R)$ to represent the observed landscape of \S\ref{Life}, then the normal gradient vector \[n_{a}=\nabla_{a}M\] (which
does not require any metric stucture for its definition) represents our perspective upon that landscape - it contains direct angular information and (in its magnitude) basic information about the local distribution of material along the uncalibrated `line of sight' normal vector.
\\
\\
Within the primitive human landscape, it is the change in perspective arising from the  act of an observer-displacement that produces the information required to assess  both 
\emph{distance traversed} and \emph{which tree nearer? which tree further?} In the case of our simple model, the change in $n_{a}$ (the perspective) arising from a displacement $dx^{k}$ can be formally
expressed as 
\begin{equation}
dn_{a}=\frac{1}{8 \pi \Sigma_F }\nabla_{i}\left(\nabla_{a}M\right)\, dx^{i}\,,\label{eqn0A}
\end{equation}
where:
\begin{itemize}
	\item  $\Sigma_F$ is a characteristic mass surface density inserted to ensure that the dimensions of $dn_a$ are the same as those of $dx^a$;
	\item the factor $8\pi$ is included for book-keeping purposes;
	\item  we assume that the connection required to give
	this latter expression an unambiguous meaning is the usual Levi-Civita
	connection - except of course, the metric tensor $g_{ab}$ required to define that connection in our Leibniz three-space is not yet defined.
\end{itemize}
Now, since $g_{ab}$ is not yet defined, then the covariant counterpart
of $dx^{a}$, given by $dx_{a}=g_{ai}dx^{i}$, is also not yet defined.
However, provided that $\nabla_{a}\nabla_{b}M$ is nonsingular,
then (\ref{eqn0A}) provides a 1:1 mapping between the contravariant
vector $dx^{a}$ and the covariant vector $dn_{a}$ so that, in the
absence of any other definition, we can \emph{define} $dn_{a}$ to
be the covariant form of $dx^{a}$. In this latter case the metric
tensor of our Leibniz three-space automatically becomes
\begin{equation}
 g_{ab}\equiv\frac{1}{8 \pi \Sigma_F }\nabla_{a}\nabla_{b}M\equiv\frac{1}{8 \pi \Sigma_F }\left(\frac{\partial^{2}M}{\partial x^{a}\partial x^{b}}-\Gamma_{ab}^{k}\frac{\partial M}{\partial x^{k}}\right),\label{(3)}
 \end{equation}
 where $\Gamma_{ab}^{k}$ are the Christoffel symbols, and given by
 \[
 \Gamma_{ab}^{k}~=~\frac{1}{2}g^{kj}\left(\frac{\partial g_{bj}}{\partial x^{a}}+\frac{\partial g_{ja}}{\partial x^{b}}-\frac{\partial g_{ab}}{\partial x^{j}}\right).\]
In this way, we arrive at a set of non-linear
differential equations defining $g_{ab}$ in terms of the unspecified mass function, $M(R)$. The
scalar product 
\begin{equation}
ds^{2}\equiv dn_{i}dx^{i}\equiv g_{ij}dx^{i}dx^{j} \label{1a}
\end{equation}
then provides the required invariant measure for the magnitude of an arbitrary infinitesimal
displacement, $dx^{a}$, in our Leibniz three-space.
\subsubsection*{Comment 1}
The crucial point of contact here with Leibniz's view of \emph{space} as a secondary construct projected out of the relationships between `objects' is simply this: the metric structure of the Leibniz three-space is defined entirely in terms of the internal properties of the primary material system, represented by $M(R)$, without reference to anything external; and  if Leibniz World is empty, that is if $M \equiv 0$, then the metric structure of the Leibniz three-space becomes wholly undefined.  
\subsubsection*{Comment 2}
Consider Mach's statement:
\begin{quote}
	\emph{... I have remained to the present day the only one who insists
		upon referring the law of inertia to the earth and, in the case of
		motions of great spatial and temporal extent, to the fixed stars ...}
	\citet{Mach1960}
\end{quote}
Given that the \lq{fixed stars}' here find their representation in $M(R)$, then there is an obvious element of this latter statement in (\ref{(3)}) above. However, any discussion of \emph{The Law of Inertia} (Newton's First Law) requires a concept of \emph{temporal evolution}, which has yet to be incorporated here. Once it has been so incoprorated, then we can imagine that Mach's statement, above, will find some form of quantitative expression in the present discussion. 
\subsection{The spherical special case }\label{sec.6} 
The system (\ref{(3)}) can be resolved by making the modelling assumption that the coordinate system is Euclidean so that $R^2 = x^i x^j\,\delta_{ij}$. 
With this understanding, it is shown, in appendix \ref{app.A}, how,
for an arbitrarily defined model of mass, $M(R)$, (\ref{(3)}) can be
exactly resolved to give an explicit form for $g_{ab}$ in terms of
such a general $M(R)$: defining the notation 
\[
\mathbf{R}\equiv(x^{1},x^{2},x^{3}),~~{\rm and}~~M'\equiv\frac{d M}{d R},
\]
this explicit form of $g_{ab}$ is given as 
\begin{equation}
g_{ab}=\frac{1}{8 \pi \Sigma_F }\left(A\delta_{ab}+Bx^{a}x^{b}\right),\label{(4)}
\end{equation}
where 
\begin{equation} 
A \equiv\frac{2\, d_0 M}{R^2},~~B\equiv-\frac{1}{2 R^2}\left(\frac{4\,d_0 M}{R^2}-\frac{M' M'}{ M}\right)   \label{4a}
\end{equation}
and $d_0$ is a dimensionless constant.
Consequently, we have the invariant line element
\begin{equation}
ds^{2}\equiv \frac{1}{8 \pi\Sigma_{F}}\left( A\, dx^{i}dx^{j}\delta_{ij} + B R^2\,dR^2\right). \label{4f}
\end{equation}
Remember that since $R \underset{\mathrm{def}}{=} G(m)$ for an undefined monotonic function $G$, then $R$ is completely uncalibrated at this juncture.
\section{The Leibniz Clock} \label{LC}
\label{sec.constr} 
So far, the concept of `time' has only entered the
discussion in a qualitative way in \S\ref{LeibnizTime} -
it has not entered in any quantitative way and, until it does, we cannot talk of \emph{velocities} or \emph{accelerations} or \emph{equations of motion}.
\\
\\
In \S\ref{LeibnizTime} we noted that, for Leibniz, \emph{time} is no more than a synonym for \emph{ordered change} within a material system. This can be rendered more usefully as the general definition that \emph{time} is a parameter which orders change within a material system. Consequently, a necessary pre-requisite for
its quantitative definition is a notion of change within such a system.
\\\\
The most simple notion of change which can be defined in a material system
is that of changing relative spatial displacements of the objects
within it. Since the systems concerned are implicitly assumed to be populated exclusively by matter with the sole  property of \emph{mass} then, in effect, all change can be described as \lq{gravitational}' change.
In existing classical theories, this fact is incorporated by constraining all
particle motions to satisfy the Weak Equivalence Principle. However, this option 
is not available in the present case, since the WEP is a dynamical principle requiring a prior
quantitative definition of \lq{time}' and such a definition is still unformulated here.  
\\
\\
This latter problem is avoided by formulating
a modified version of the WEP which notes that the geometric shapes of gravitational trajectories in ordinary physical space are themselves independent of the internal properties
of the particles concerned. So we arrive at the principle:
\begin{quote}
	\textbf{\emph{Shape Independence Principle:}} \emph{When a massive test-particle moves under the influence of \lq{gravitation}' only, the shape of that particle's trajectory in ordinary geometric three-space is independent of the intrinsic properties of the particle concerned.}  
\end{quote}
\subsection{Linear dependence of the Euler-Lagrange equations}\label{EoM}
Suppose $p$ and $q$ are two arbitrarily chosen point coordinates
on the trajectory of the chosen particle, and suppose that (\ref{1a})
is used to give the scalar invariant 
\begin{equation}
I(p,q)=\int_{p}^{q}\sqrt{dn_{i}dx^{i}}\equiv\int_{p}^{q}\sqrt{g_{ij}dx^{i}dx^{j}}. \label{(2)} 
\end{equation}
Then,  $I(p,q)$
gives a scalar record of how the particle has moved between $p$ and
$q$ defined with respect to the particle's continually changing relationship
with the  mass model, $M(R)$.\\
\\
Now suppose $I(p,q)$ is minimized with respect to choice of the
trajectory connecting $p$ and $q$, then this minimizing trajectory,
and its shape in particular,
is independent of the internal properties of the particle concerned 
so that the \emph{Shape Independence Principle} above is satisfied. 
\\
\\
Consequently, defining the Lagrangian density in the usual way, and using (\ref{4f}), we have
\begin{eqnarray}
I(p,q)&=&\int^q_p{\cal L}\,dt\equiv \int^q_p\sqrt{g_{ij}\dot{x}^{i}\dot{x}^{j}}\,dt 
\nonumber \\
 &=& \left(\frac{1}{8 \pi\Sigma_{F}}\right)^{1/2}\int^q_p\left( A \dot{x}^{i} \dot{x}^{j}\delta_{ij} + B R^2\, \dot{R}^2\right)^{1/2}\,dt  \label{4d}
\end{eqnarray}
where $A$ and $B$ are defined at (\ref{4a})
and $t$ is a temporal ordering parameter. At this stage, we note that $\mathcal{I}(p,q)$ is homogeneous degree zero in $t$ which means that it is invariant under $t \rightarrow f(t)$ for any monotonic function $f$. Consequently,  $t$ cannot be uniquely identified as physical time.
\\\\
Furthermore, since the system $\mathcal{I}(p,q)$ is homogeneous degree zero in $t$ then,  by a standard result, the Euler-Lagrange (E-L) equations for the system  are not linearly independent. It follows that additional information is required to close the system, and that this additional information will amount to defining  what is meant by \lq{physical time}'. In practice, as will be shown, the requirement that the system is conservative leads to the  \emph{geometric Leibniz Clock} being defined according to:
\[
dt^2 = \left(\frac{1}{k^2 A^2}\right) g_{ij} {dx}^{i} {dx}^{j}
\] 
for some constant $k$ for the particular case of non-circular motions.
\\\\ 
An identical circumstance arises in General Relativity when the equations
of motion are derived from an action integral which is formally identical
to (\ref{(2)}). In that case, the system is closed by specifying
the arbitrary parameter involved to be {\lq particle proper time'} defined according to:
\begin{equation}
d\tau^2 = \left(\frac{1}{c^2}\right) g_{ij} {dx}^{i} {dx}^{j},\label{(5a)}
\end{equation}
where $c$ is the usual light speed. 
\subsection{The Leibniz Clock for non-circular motions}\label{General-state}
In the following, we assume that the E-L equations are expressed in spherical polars, and that all motions occur in the equatorial plane, $\phi=\pi/2$.  
For the non-degenerate state of strictly non-circular motions, it can be shown that the E-L equations can be combined to give the equations of motion in standard vector form as:  
\begin{eqnarray}
\mathbf{\ddot{R}} &+&  \alpha \, \mathbf{\dot{R}}  
+ \beta\,\mathbf{R}
= 0, \label{(5)} \\
\alpha &\equiv& \frac{1}{2A} \left(2A'\dot{R}-2\frac{\dot{{\cal L}}}{{\cal L}}A\right) \label{5B}\\
\beta &\equiv& \frac{1}{2A} \left(B' R \dot{R}^{2}+2B\left(\dot{R}^2+R \ddot{R} \right)-\frac{A'}{R}\left(\mathbf{\dot{R}\cdot\dot{R}}\right)-2\frac{\dot{{\cal L}}}{{\cal L}}B R \dot{R}\right),
\label{5C}
\end{eqnarray}
where $A$ and $B$ are defined at (\ref{4a}).
For the reasons stated in \S\ref{EoM}, the individual component equations of this system cannot be linearly independent. Consequently, there is a need for additional information which will, in effect, \emph{define} physical time - the geometric Leibniz Clock - for the system.
\\\\
To this end, it is clear from (\ref{(5)}) that if this additional information takes the form of the requirement that the system must conservative, then the condition $\alpha = 0$ must necessarily be imposed so that dissipation is eliminated. But in this case, (\ref{(5)}) becomes $\mathbf{\ddot{R}} + \beta\,\mathbf{R}= 0$ which
decomposes into a non-trivial radial equation together with the standard equation for angular momentum conservation and these two equations \emph{are} linearly independent, and therefore have unique solutions (to within initial conditions).  
\\\\
Consequently, the only way the condition $\alpha=0$ can be consistent with these unique solutions (which it must be) is if it is actually a solution of the radial component of (\ref{(5)}) in the first place. It is shown in \S\ref{LeibnizClocks} that this is actually the case here so that $\alpha=0$ defines the geometric Leibniz Clock for the system.
\\\\
Finally, it follows that the total independent content of (\ref{(5)}) is that every particle is an energy and angular momentum conserving geometric Leibniz Clock.
\subsection{The potential form of the Leibniz Clock:} \label{PhysicalTime}
Given the Leibniz Clock defined by $\alpha=0$, then (\ref{5B}) yields
\[
\alpha \equiv \frac{1}{2A} \left(2A'\dot{R}-2\frac{\dot{{\cal L}}}{{\cal L}}A\right) = 0.
\]
Consequently, since $\dot{R}\neq 0$ then the Leibniz Clock can be expressed as:
\begin{equation}
\frac{A'}{A}\dot{R}=\frac{\dot{{\cal L}}}{{\cal L}}~~\rightarrow~~{\cal L}=  \frac{v_0}{8 \pi \Sigma_F} A, \label{(6)}
\end{equation}
where $v_0$ is an undetermined parameter with units of \emph{velocity}  and $\Sigma_F$ is  the characteristic mass surface density parameter introduced at equation (\ref{eqn0A}), and the factor $8\pi$ is introduced for book-keeping purposes. From (\ref{4d}) and (\ref{(6)}) we now get
\begin{equation}
{\cal L}^2 \equiv g_{ij}\dot{x}^{i}\dot{x}^{j}  = \frac{v_0^2}{64  \pi^2 \Sigma_F ^2} A^{2}\label{(7)}
\end{equation}
so that the  elapsed Leibniz Clock-time arising from a spatial displacement is \textit{defined} by:
\[
dt^{2}\, \underset{\mathrm{def}}{=} \, \frac{1}{v_0^2}\left(\frac{ 64 \pi^2 \Sigma_F ^2}{ A^{2}}\right)g_{ij}dx^{i}dx^{j} \rightarrow \]
\begin{equation}
dt^{2}\, \underset{\mathrm{def}}{=}  \frac{1}{v_0^2 } \left(\frac{8\pi \Sigma_F }{ A^2}\right)\left[ A\,\left(dR^2 + R^2 d\theta^2 \right)+ B R^2 dR^2\right],\label{(8)}
\end{equation}
from which we now see that $v_0$, which has dimensions of \emph{velocity}, is, in fact, a conversion factor between units of \emph{distance} and units of \emph{time}, which is precisely the interpretation that Bondi attached to the light-speed constant, $c$, in the 1960s. We shall refer to $v_0$ as the \emph{clock-rate parameter} and, for now, leave its value unassigned.   
\\
\\
In any event, according to the above, the elapsing of internal time for the system concerned is given a direct physical interpretation
in terms of the process of \textit{spatial displacement} within that system. Thus, just as we have shown that \lq{metrical space}' can be considered to be projected out of the relationships within the material world, so \lq{elapsed Leibniz Clock-time}' can be considered to be projected out of \lq{process}' within that world, which conforms exactly with the Leibniz view on the nature of \lq{time}' expressed in \S\ref{LeibnizTime}.
\\
\\
With this understanding, (\ref{(8)}) can be written as
\begin{equation}
\dot{R}^2 + R^2 \dot{\theta}^2  \,=\, \frac{ v_0^2}{ 8 \pi \Sigma_F } A - \frac{B}{A} R^2 \dot{R}^2  \label{ClockConstraint}
\end{equation}
which, following the algebra of \S\ref{C} to eliminate the right-hand-side time derivative, gives the potential form of the Leibniz clock:
\begin{equation}
\frac{1}{2}\left( \dot{R}^2 + R^2 \dot{\theta}^2 \right) = \frac{1}{2}\left( 4 \,d_0^2 v_0^2\left(\frac{  M}{4 \pi \Sigma_F } -\frac{ h^2 }{d_0 v_0^2} \right)\left( \frac{  \,M^2}{R^4 M' M'} \right)+\frac{h^2}{R^2} \right)\label{(8A)}
\end{equation}
where $h$ is conserved angular momentum.
\subsection{The Leibniz Clock solves the radial equation of (\ref{(5)})} \label{LeibnizClocks}
By the considerations of \S\ref{General-state}, if the Leibniz Clock $\alpha=0$ satisfies the radial component of
\begin{eqnarray}
\mathbf{\ddot{R}} &+& \beta \,\mathbf{R} = 0, \label{(9)} \\
\beta &\equiv& \frac{1}{2A} \left(B' R \dot{R}^{2}+2B\left(\dot{R}^2+R \ddot{R} \right)-\frac{A'}{R}\left(\mathbf{\dot{R}\cdot\dot{R}}\right)-2\frac{A'}{A}B R \dot{R}^{2}\right) \nonumber
\end{eqnarray}
then it must also satisfy the radial component of (\ref{(5)}).
Following appendix \S\ref{app.C}, in an analysis which makes explicit use of the condition $R\neq constant$, we find that this latter system can be written in potential form as
\begin{eqnarray}
\mathbf{\ddot{R}} &=& - \frac{d\mathcal{V}}{dR}\,\mathbf{\hat{R}},
\label{11R} \\
\mathcal{V} &\equiv&  -\frac{1}{2}\left(4 \,d_0^2 v_0^2\left(\frac{  M}{4 \pi \Sigma_F } -\frac{ h^2 }{d_0 v_0^2} \right)\left( \frac{  \,M^2}{R^4 M' M'} \right)+\frac{h^2}{R^2} \right).
\label{11S}
\end{eqnarray}
The radial component of (\ref{11R}) can be trivially integrated to give
\begin{equation}
\frac{1}{2}\left( \dot{R}^2 + R^2 \dot{\theta}^2 \right) \,=\,- \mathcal{V} \label{11T}
\end{equation}
which, given the definition of $\mathcal{V}$ at (\ref{11S}), is identical to the potential form of the Leibniz Clock given at (\ref{(8A)}) for the result. 
\\\\
In conclusion, the net content of (\ref{(5)}) is that every particle is an energy and angular momentum conserving Leibniz Clock.
\subsection{A Leibniz Clock is synchronous to a classical standard clock}
In the following we show, as a trivial mathematical result, that a Leibniz clock defined by  (\ref{(8A)}) is entirely equivalent to a classically defined standard clock, with each exactly synchronized to the other.
\subsubsection*{The Leibniz point of view}
The Leibniz Clock in the form of (\ref{11T}) can be simplified and rearranged to define the elapsed time $dt$ corresponding to a radial displacement $dR$ as: 
\begin{equation}
dt \underset{\mathrm{def}}{=}  \left[ 4 \,d_0^2 v_0^2\left(\frac{  M}{4 \pi \Sigma_F } -\frac{ h^2 }{d_0 v_0^2} \right)\left( \frac{  \,M^2}{R^4 M' M'} \right) \right]^{-1/2}\,dR. \label{Clock1}
\end{equation}
\subsubsection*{The classical point of view}
From the classical perspective, we interpret the parameter $t$ occurring in the equation of motion of (\ref{11R}) as the \emph{clock time} measured by an external classical clock, ticking off standard units of time, \emph{seconds, minutes, hours} etc. The Leibniz clock of (\ref{11T}) is correspondingly re-interpreted as the classical \emph{energy equation} arising from the first integration of the equation of motion wrt classical clock-time.
This \emph{energy equation} can now be rearranged to give the spatial displacement corresponding to a given elapsed classical clock-time interval as:
\begin{equation}
dR = \left[ 4 \,d_0^2 v_0^2\left(\frac{  M}{4 \pi \Sigma_F } -\frac{ h^2 }{d_0 v_0^2} \right)\left( \frac{  \,M^2}{R^4 M' M'} \right) \right]^{1/2}\,dt. \label{Clock2}
\end{equation}
Note that, unlike (\ref{Clock1}), this is not a definition but is rather the application of a classical equation of motion to determine a spatial displacement in terms of a time-interval determined by an external classical clock.
\subsubsection*{Conclusions}
It is trivially true that (\ref{Clock1}) and (\ref{Clock2}) are, in formal terms, simple rearrangements of each other so that if a given $(dR, dt)$ satisfies either one of these expressions, then it must necessarily satisfy the other. In other words the Leibniz Clock is synchronous to any external classical standard clock used to measure the passage of classical time for the system. 
\subsubsection*{Comment}
Whilst this result is mathematically trivial, it is a non-trivial result from the point of view of physics for it is an exact formal realization of the Leibniz assertion that the notion of \emph{time passing} is no more, and no less, than an idealized model for the process of \emph{ordered change} within a material system.  From this point of view, classical clock-time (that is, \emph{external time}) is an un-necessary abstraction and, arguably, is an idea which has led gravitational physics in particular into the conceptual \emph{cul-de-sac} of the spacetime continuum.
\subsection{The assignment of value to the clock-rate parameter $v_0$} \label{v_0}
As soon as we re-interpret (\ref{11R}) as a classical particle equation of motion (as above), then it becomes clear how the value of the clock-rate parameter, $v_0$, must be assigned: specifically, when the classical interpretation is employed, the Weak Equivalence Principle (WEP) becomes applicable, which means that particle accelerations must be independent of any properties that the particle might possess. It follows that, for any given particle, the value of $v_0$ (which has units of \emph{velocity}) must be assigned accordingly. There are then two cases for (\ref{11R}) on the classical interpretation:
\begin{itemize}
	\item $\mathcal{V} = constant$, so that particle accelerations are identically zero implying, in turn, that the WEP imposes no constraints on the parameter $v_0$. For this case, (\ref{11T}) becomes
	\[
	\dot{R}^2 + R^2 \dot{\theta}^2  \,=\, v_0^2
	\]  
	which, on the classical interpretation of (\ref{11R}), is the particle's energy equation implying that $v_0$  must be identified with the velocity magnitude of the particle concerned;
	\item $\mathcal{V} \neq constant$, so that particle accelerations are non-trivial. The WEP can then only be satisfied if the parameter $v_0$ is independent of the particle motion. This implies that $v_0$ must be a  constant of the overall system of which the particle is a part. So, for example, it could be the velocity dispersion for the system.
\end{itemize} 
However, whilst the classical interpretation of (\ref{11R}) has indicated how the value of $v_0$ should be assigned in each of the two situations, we must remember that, within the geometric Leibniz synthesis,  $v_0$ is the clock-rate parameter which acts as a conversion factor between units of distance and units of time, and is not a velocity magnitude at all. 
\\\\
It follows that in the case of $\mathcal{V}=constant$, assigning $v_0$ values according to particle velocity magnitudes on the classical interpretation is equivalent to assigning $v_0$ values according to the condition that all Leibniz Clocks are synchronized to \emph{tick}  at the rate of the classical clock used in the classical interpretation.
\section{The equilibrium state}\label{SimpleModel}
The work of \citet{Baryshev1995}  and many others (see appendix \S\ref{FractalUniverse} for an overview)  has shown how the distribution of galaxies on medium scales ($100 \sim 300\,Mpc$) is quasi-fractal, $D\approx 2$.  If the idea of a quasi-fractal universe on medium scales is idealized to include all physical scales then  we can say that, \emph{about any point} chosen as the centre, mass is distributed according to
\begin{equation}
\mathcal{M}(R) = 4 \pi R^2 \,\Sigma_F; \,\,\,R < \infty, \label{Frac1}
\end{equation}
where $\Sigma_F$ is the characteristic mass surface density of the distribution. Whilst it is straightforward to show that (\ref{11R}) gives $\mathbf{\ddot{R}}=0$ with this mass distribution - so that a state of dynamical equilibrium exists - it is informative to use the classical argument to make the same point.
\\\\
So, since the idealized matter distribution of (\ref{Frac1}) is isotropic (by definition) about any arbitrarily chosen centre, then the notional gravitational acceleration imparted to a particle at radius $R$, and generated by the material contained within $R$, is directed towards the chosen centre and has magnitude given by
\[
a_F \equiv \frac{\mathcal{M}(R) \, G}{R^2} = 4 \pi G\, \Sigma_F;\,\,\, R< \infty
\]
so that $a_F$ can be interpreted as the characteristic acceleration scale associated with a $D=2$ fractal distribution of material of mass surface density $\Sigma_F$.
\\\\
On this basis, it is clear that the \emph{net} actual gravitational acceleration imparted to a material particle immersed anywhere in the global distribution (\ref{Frac1}) is zero, from which it can be concluded that a $D=2$ fractal distribution of material is in a state of dynamical equilibrium.
\\\\
It follows that:
\begin{itemize}
	\item if a finite spherical volume, radius $R_0$, is imagined emptied of all material, then the net actual gravitational acceleration of any material particle placed on $R_0$  will be $a_F$ directed radially outwards from the centre of the empty volume;
	\item the empty spherical volume is unstable since all accelerations on $R_0$ are outward. It follows that stability requires the volume to be occupied by a stablizing mass, a galaxy say, creating a state of zero net radial acceleration on $R_0$. In other words, the equilibrium condition
	\begin{equation}
	\frac{V_0^2}{R_0} = a_F \label{eqn2HH}
	\end{equation}
	is required.
\end{itemize}
\section{The roots of MOND in a Leibniz Clock} \label{OverView}
We show how a geometric Leibniz Clock for a simple spherical galaxy model provides a detailed understanding of Milgrom's MOND algorithm.
\subsection{Overview} \label{OverView1}
We hypothesize that the $D \approx 2$ quasi-fractal equilibrium state extends to the finer scales of the intergalactic medium and assume that any isolated galaxy is in a state of dynamical equilibrium within that IGM. We model such a galaxy as a finite bounded spherical perturbation of the IGM, and then write down the Leibniz Clock for this model galaxy assuming purely circular motions. This model galaxy then has the following irreducible properties:
\begin{itemize}
	\item The rotation curve is necessarily asymptotically flat;
	\item The assumed state of dynamical equilibrium requires that the gravitational acceleration on the perturbation boundary, $R=R_0$, generated by the interior mass is balanced by the gravitational acceleration generated by the exterior mass of the IGM. Consequently, by (\ref{eqn2HH}), the equilibrium condition $V_0^2/R_0 = a_F$ is required;
	\item The baryonic Tully-Fisher relationship in the form
	\[
	V_{flat}^4 = a_F \,G\,M_{flat}~~{\rm where}~~M_{flat} \equiv \left(\frac{V_{flat}}{V_0} \right)^2 M_0
	\] 
	and where $M_0$ is the total of baryonic mass contained within the perturbation boundary, $R = R_0$, is necessarily satisfied;
	\item Using the SPARC sample of \citet{Lelli2016A} together with the photometric mass-models of \citet{Lelli2016B} used to determine $M_0$, a direct computation based upon invariance properties of the isolated galaxy model yields $a_F \approx 1.2 \times 10^{-10}\,mtrs/sec^2$ as the characteristic acceleration scale of the $\Sigma_F$ IGM.
\end{itemize}
It is obvious from these considerations that the characteristic acceleration scale $a_F$ and MOND's critical acceleration parameter, $a_0$, are one and the same thing. 
Apart from the difference of interpretation around the requirement $V_0^2/R_0 = a_F$, the quantitative definition of $M_{flat}$ via a scaling law, and the fact that the quantitative evaluation of $a_F$ is a prediction arising from a direct computation based on model invariance properties, this list of irreducible properties for the model galaxy characterizes completely and exactly those galaxy properties that MOND either requires (flatness), assumes (a critical acceleration scale) or predicts (the baryonic Tully-Fisher relation). 
\\\\
In other words, MOND works as well as it does because its requirements, assumptions and predictions match the reality of the isolated galaxy described by the geometric Leibniz Clock.
\subsection{The Leibniz Clock for circular motions}\label{Degenerate-state}
Whilst the Leibniz clock for the case of non-circular motions has been considered in great detail in \S\ref{LC}, the roots of MOND are actually to be found in the Leibniz clock defined purely for the case of strictly \emph{circular motions} for which $\dot{R} = 0$ by definition. Using the Lagrangian density $\mathcal{L}$ defined at (\ref{4d}), the degenerate radial E-L equation, ${\partial {\cal L}}/{\partial R} = 0$,  is then trivially integrated to give the Leibniz Clock for this case as:
\begin{equation}
{\cal L}^2 = v_0^2 ~~\rightarrow~~ R^2\dot{\theta}^2 = \frac{8\pi\Sigma_Fv_0^2}{A} \equiv \frac{v_0^2}{d_0}\,\left(\frac{4\pi\Sigma_FR^2}{M} \right) \equiv  \frac{v_0^2}{d_0}\,\left(\frac{\Sigma_F}{\Sigma_R} \right)\label{(11)}
\end{equation}
where $\Sigma_R$ is the mass surface density at radius $R$ and $v_0$ is a constant having dimensions of velocity. The transverse equation of motion gives angular momentum conservation, trivially. 
\subsection{The spherical Leibniz galaxy model: details}
The irreducible basic expression of geometric Leibniz cosmology is a world of dynamic equilibrium within which material is necessarily non-trivially present in the form of a $D=2$ fractal distribution so that, about any point chosen as the centre, mass is distributed according to
\begin{equation}
M(R) = 4 \pi \Sigma_F R^2  \label{Frac1-A}
\end{equation}
constrained to be valid about any centre, where $\Sigma_F$ is the characteristic mass surface density scale and $a_F \equiv 4\pi G\,\Sigma_F$ is the associated characteristic acceleration scale. 
\\\\
We now make the hypothesis that this  $D=2$ fractal equilibrium state extends to the finer scales of the intergalactic medium (IGM) itself, and that an isolated galactic object can be modelled as a finite bounded perturbation of this idealized IGM within the specific context of the geometric Leibnizian worldview.
\\\\
Given the foregoing, and noting that Milgrom's idea of a critical acceleration boundary around every isolated galactic object amounts to the idea of a finite boundary around that object, then we model an isolated galaxy as a finite bounded spherical perturbation of (\ref{Frac1-A}). It therefore has the general form:
\begin{eqnarray}
M(R) &\equiv& M_g(R),~~~~ R \leq R_0; \label{Frac2-A} \\
M(R) &\equiv&  M_g(R_0)  + 4 \pi (R^2-R_0^2)\,\Sigma_F,~~~~ R > R_0, \nonumber
\end{eqnarray}
where $M_g(R)$ represents the baryonic galactic mass contained within radius $R\leq R_0$ and $R_0$ is the finite, but otherwise unspecified, radial boundary of the perturbation. 
\\\\
Consequently, with this isolated galaxy model, and writing $v_0^2/d_0 \equiv V^2_{flat}$, the Leibniz Clock for the case of purely circular orbits given at (\ref{(11)})  becomes  the scaling relationship:
\begin{eqnarray}
\left[\frac{V_{rot}(R)}{V_{flat}}\right]^2 \Sigma_R &=& \Sigma_F, ~~~~ R < \infty; \label{eqn2} \\  
\Sigma_R &\equiv& \frac{M_g(R) }{4 \pi R^2},~~~~ R \leq R_0; \nonumber \\
\Sigma_R &\equiv& \frac{M_g(R_0)  + 4 \pi (R^2-R_0^2)\,\Sigma_F}{4 \pi R^2},~~~~ R > R_0 \nonumber
\end{eqnarray}
from which it is clear that the rotation curve is automatically asymptotically flat.
Additionally, for the modelled galaxy to be in a state of dynamical equilibrium as it must be then, according to (\ref{eqn2HH}), the equilibrium condition \[\frac{V_0^2}{R_0} = a_F\] 
must be satisfied.
\\\\
Finally, to see that the baryonic Tully-Fisher relationship is satisifed, evaluate (\ref{eqn2}) at $R=R_0$, eliminate $R_0$ using the equilibrium condition above, and then use $\Sigma_F= a_F/(4\pi G)$ to get:
\begin{equation}
V_{flat}^4 = a_F\, G\, M_{flat}{\rm(theory)}  \label{eqn5C}
\end{equation}
where
\[M_{flat}{\rm(theory)} \equiv \left(\frac{V_{flat}}{V_0} \right)^2 M_g(R_0)\]
and $M_g(R_0)$ is the baryonic galactic mass contained within the perturbation boundary, $R_0$. In practice, $M_g(R_0)$ is estimated photometrically using the SPARC sample of \citet{Lelli2016A} together with the mass-models \citet{Lelli2016B}.
\\\\
Given that $M_{flat}{\rm{(theory)}}$ above is in an almost perfect statistical correspondence with photometric determinations of $M_{flat}$ - which it is found to be, \citet{Roscoe2022} - then (\ref{eqn5C}) is identical to Milgrom's form of the baryonic Tully-Fisher relationship, but with $a_0$ replaced by $a_F$.
\subsection{Practical evaluation of $(a_F,\,\Sigma_F)$} \label{Gamma1}
\begin{figure}[H]
	\centering
	\includegraphics[width=0.7\linewidth]{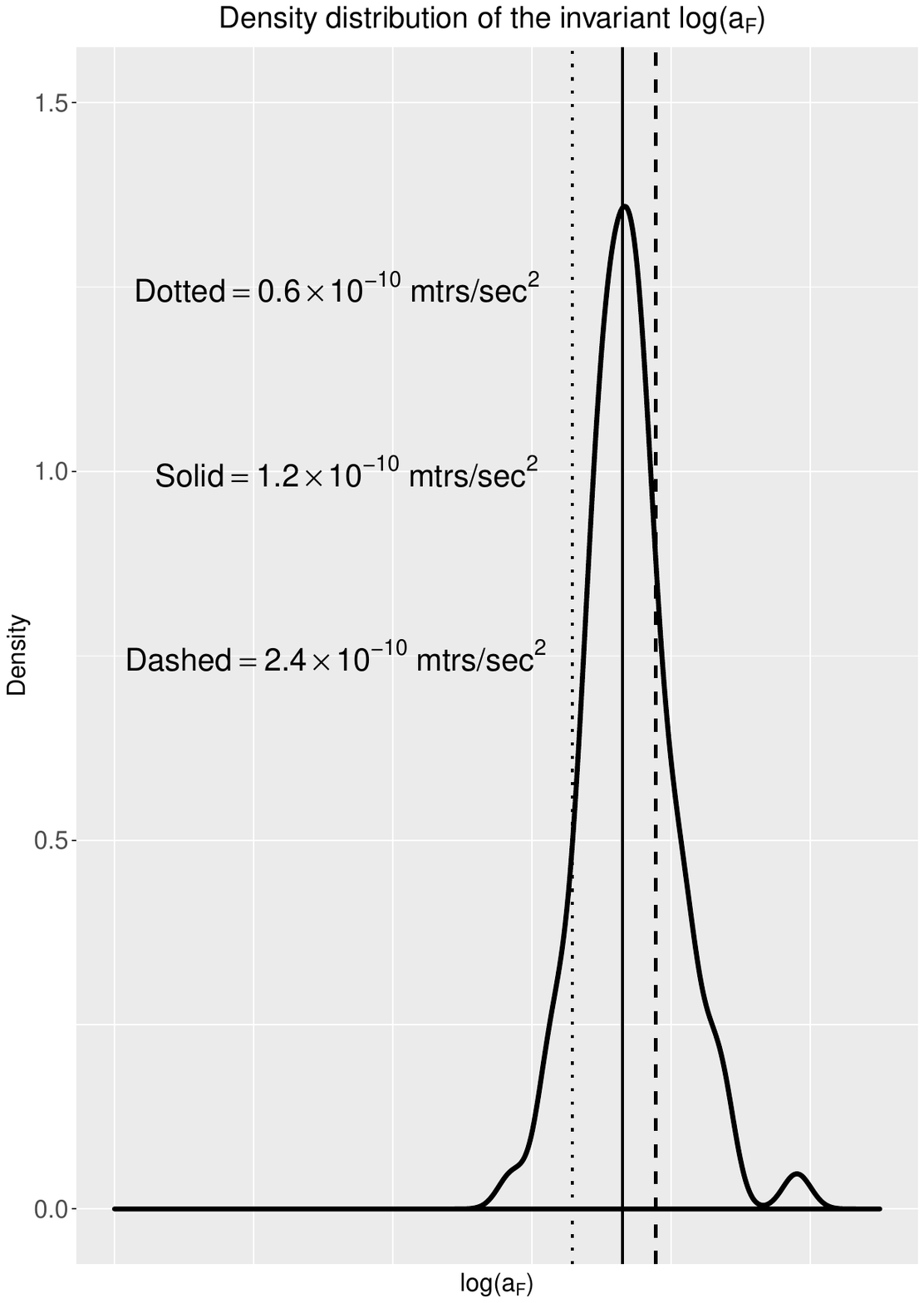}
%	\includegraphics[width=0.7\linewidth]{DensityGamma}
%	\caption{Solid black curve = density distribution of $\log(\Gamma)$. Here, $\Upsilon_* = 0.8$. }
	\caption{Solid black curve = density distribution of $\log(a_F)$. Here, $\Upsilon_* = 0.8$. }
	\label{GammaDensity}
\end{figure}

Whilst the full details of the evaluation are given in \citet{Roscoe2022}, a brief outline is given in the following.
\\\\
The basic resource is the SPARC sample of \citet{Lelli2016A}.
Following \citet{Lelli2016B}, all required baryonic mass determinations in galaxy disks are then provided by the model
\begin{equation}
M_{bar} = M_{gas} + \Upsilon_*\, L_{[3.6]}, \label{MassModel1}
\end{equation} 
where $M_{gas}$ is the gas mass, $L_{[3.6]}$ is the $[3.6]$ luminosity and $\Upsilon_*$ is the stellar MLR, assumed here to be equal for the disk and the bulge (where present), and constant across the whole SPARC sample.
Lelli et al find that the optimal choice for $\Upsilon_*$ (which minimizes scatter in the baryonic Tully-Fisher relationship) satisfies $\Upsilon_* \geq 0.5$. For current purposes we restrict the choice to $\Upsilon_* \in (0.5,\,1.0)$.
\\\\
In practice, because of uncertainties in estimating the absolute position of the perturbation boundary, $R_0$, on the rotation curve then, across the whole sample with a globally applied stellar MLR, $\Upsilon_* \in (0.5,\,1.0)$, we get distributions of values for $a_F$ similar to that plotted in figure \ref{GammaDensity} for which $\Upsilon_*=0.8$. These distributions are sharply modal and, using the modal values of these distributions as the estimate for $a_F$ in each case, we find:
\[
a_F \in (1.4,\,0.9) \times 10^{-10}\,mtrs/sec^2. 
\]
In particular, the choice $\Upsilon_* = 0.8$ yields $a_F \approx 1.2 \times 10^{-10}\,mtrs/sec^2$ which corresponds to $\Sigma_F \approx 0.14\, kg/mtrs^2$ for the mass surface density of the fractal IGM.
\section{Summary and Conclusions}
In modern terms, Leibniz viewed \emph{space} as a secondary construct projected out of primary relationships between objects whilst he considered the concept of \emph{time} to be no more than a synonym for \emph{ordered process}. 
\\\\
From the point of view of quantifying these ideas, it is arguable that \emph{space} presents the greatest difficulty simply because  if it is a secondary construct projected out of primary relationships between objects then, in fact, it stands as a proxy for its own contents - they are one and the same thing, in effect. This implies a deeply recursive relationship, and therein lies the main difficulty associated with any attempt to synthesize a quantitative Leibnizian worldview.
\\\\
The arguments presented here rest primarily upon interpreting this recursive view of \lq{space and its contents}' to imply the weaker statement: \emph{without material content no concept of metrical space can exist} which, in \S\ref{Leibniz}, is mutated into a quantitative theory of Leibnizian metrical space. Subsequently, it is shown in \S\ref{LC} how the concept of \emph{time}, considered as a synonym for \emph{ordered process}, becomes almost seamlessly incorporated into this realization of the Leibnizian metrical space, leading immediately to the result that any gravitating particle, or system of particles, is no more than an energy and angular-momentum conserving clock - the geometric Leibniz Clock of the title.
\\\\
The whole system has an irreducible equilibrium state which consists of a global $D=2$ fractal distribution of material which has a characteristic mass surface density $\Sigma_F$ with an associated characteristic acceleration $a_F\equiv 4 \pi G \Sigma_F$. When we model an isolated galactic object, in equilibrium with its environment, as a finite bounded spherically symmetric perturbation, radius $R_0$, of this equilibrium state and consider the special case of purely circular motions within this system,  we find that the object is characterized by the properties:
\begin{itemize}
	\item the rotation curve is asymptotically flat;
	\item there is an equilibrium condition $V_0^2/R_0 = a_F$ which must be satisfied on $R_0$;
	\item the baryonic Tully-Fisher relationship in the form $V_{flat}^4 = a_F G M_{flat}$ is satisfied. 
\end{itemize}
Discounting the classical wrapping of MOND (its use of the Newtonian force concept), it is clear that the isolated MOND galaxy is structurally identical to the isolated Leibniz galaxy. This conclusion is unambiguously consolidated when a direct calculation, based upon the invariance properties of the geometric Leibniz Clock for the system, and using the SPARC data of \citet{Lelli2016A} and the photometric mass-models of \citet{Lelli2016B}, gives $a_F \approx 1.2 \times 10^{-10}\,mtrs/sec^2$.

%%%%%%%%%%%%%%%%%%%%%%%%%%%%%%%%%%%%%%%%%%%%%%%%%%

%%%%%%%%%%%%%%%%% APPENDICES %%%%%%%%%%%%%%%%%%%%%

\appendix
\section{Medium to large scale structure}\label{FractalUniverse}
A basic assumption of the \textit{Standard Model}
of modern cosmology is that, on some scale, the universe is homogeneous;
however, in early responses to suspicions that the accruing data was
more consistent with Charlier's conceptions \citet{Charlier1908, Charlier1922, Charlier1924}
of an hierarchical universe than with the requirements of the \textit{Standard
	Model},  \citet{De Vaucouleurs1970} showed that, within wide limits,
the available data satisfied a mass distribution law $M\approx r^{1.3}$,
whilst \citet{Peebles1980} found $M\approx r^{1.23}$. 
\subsection{Modern observations and the debate}\label{Observations}
The situation,
from the point of view of the \textit{Standard Model}, continued to
deteriorate with the growth of the data-base to the point that, \citet{Baryshev1995}  were able to say
\begin{quote}
	\emph{...the scale of the largest inhomogeneities (discovered to date)
		is comparable with the extent of the surveys, so that the largest
		known structures are limited by the boundaries of the survey in which
		they are detected.}  
\end{quote}
For example, several redshift surveys of the late 20th century, such
as those performed by \citet{Huchra1983}, \citet{Giovanelli1986}, \citet{DeLapparent1988}, \citet{Broadhurst}, \citet{DaCosta} and \citet{Vettolani} 
etc discovered massive structures such as sheets, filaments, superclusters
and voids, and showed that large structures are common features of
the observable universe; the most significant conclusion drawn from
all of these surveys was that the scale of the largest inhomogeneities
observed in the samples was comparable with the spatial extent of
those surveys themselves.\\
\\
In the closing years of the century, several quantitative analyses
of both pencil-beam and wide-angle surveys of galaxy distributions
were performed: three examples are given by \citet{Joyce}  who analysed the CfA2-South catalogue to find fractal
behaviour with $D\,$=$\,1.9\pm0.1$; \citet{SylosLabini}
analysed the APM-Stromlo survey to find fractal behaviour with $D\,$=$\,2.1\pm0.1$,
whilst \citet{SylosLabini1} analysed the
Perseus-Pisces survey to find fractal behaviour with $D\,$=$\,2.0\pm0.1$.
There are many other papers of this nature, and of the same period,
in the literature all supporting the view that, out to $30-40h^{-1}Mpc$
at least, galaxy distributions appeared to be consistent with the simple stochastic fractal model with the critical fractal dimension of $D\approx  D_{crit} = 2$.\\
\\
This latter view became widely accepted (for example, see  \citet{Wu}), and the open question became whether or not
there was transition to homogeneity on some sufficiently large scale.
For example, \citet{Scaramella} analyse the ESO Slice
Project redshift survey, whilst \citet{Martinez} analyse
the Perseus-Pisces, the APM-Stromlo and the 1.2-Jy IRAS redshift surveys,
with both groups claiming to find evidence for a cross-over to homogeneity
at large scales.\\
\\
At around about this time, the argument reduced to a question of
statistics (\citet{Labini}, \citet{Gabrielli}, \citet{Pietronero}):
basically, the proponents of the fractal view began to argue that
the statistical tools (that is, two-point correlation function methods) widely used
to analyse galaxy distributions by the proponents of the opposite
view are deeply rooted in classical ideas of statistics and implicitly
assume that the distributions from which samples are drawn are homogeneous
in the first place.  \citet{Hogg}, having accepted
these arguments, applied the techniques argued for by the pro-fractal
community (which use the \emph{conditional density} as an appropriate
statistic) to a sample drawn from Release Four of the Sloan Digital
Sky Survey. They claimed that the application of these methods does
show a turnover to homogeneity at the largest scales thereby closing,
as they see it, the argument. In response, \citet{SylosLabini2} 
criticized their paper on the basis that the strength of the
conclusions drawn is unwarrented given the deficencies of the sample
- in effect, that it is not big enough. More recently, \citet{Tekhanovich} have addressed the deficencies of the Hogg et al analysis by analysing the 2MRS catalogue, which provides redshifts of over 43,000 objects out to about 300Mpc, using conditional density methods; their analysis shows that the distribution of objects in the 2MRS catalogue is consistent with the simple stochastic fractal model with the critical fractal dimension of $D\approx  D_{crit} = 2$.  \\
\\
To summarize, the proponents of non-trivially fractal large-scale
structure have won the argument out to medium distances and the controversy
now revolves around the largest scales encompassed by the SDSS. 
\section{A Resolution of the Metric Tensor}\label{app.A} 
The algebra of this section is most easily performed by the change of notation:
\[
\Phi \equiv \frac{R^2}{2},~~~~ R^{2}\equiv\left(x^{1}\right)^{2}+\left(x^{2}\right)^{2}+\left(x^{3}\right)^{2},~~~
\]
\[
 M' \equiv \frac{d M}{d \Phi} \equiv \frac{1}{R}\frac{d M}{dR},~~~~M'' \equiv \frac{d^2 M }{d\Phi^2},~~~{\rm etc}.
\]
The general system is given by 
\[
g_{ab}=\frac{\partial^{2}M}{\partial x^{a}\partial x^{b}}-\Gamma_{ab}^{k}\frac{\partial M}{\partial x^{k}},
\]
\[
\Gamma_{ab}^{k}~\equiv~\frac{1}{2}g^{kj}\left(\frac{\partial g_{bj}}{\partial x^{a}}+\frac{\partial g_{ja}}{\partial x^{b}}-\frac{\partial g_{ab}}{\partial x^{j}}\right),\]
and the first major problem is to express $g_{ab}$ in terms of the
mass function, $M$. The key to resolving this is to note the relationship
\[
\frac{\partial^{2}M}{\partial x^{a}\partial x^{b}}=M'\delta_{ab}+M''x^{a}x^{b},
\]
where $M'\equiv d M/d\Phi$, $M''\equiv d^{2}M/d\Phi^{2}$, $\Phi\equiv R^{2}/2$,
since this immediately suggests the general structure 
\begin{equation}
g_{ab}=A\delta_{ab}+Bx^{a}x^{b},\label{A1}
\end{equation}
for unknown functions, $A$ and $B$. It is easily found that \[
g^{ab}={\frac{1}{A}}\left[\delta_{ab}-\left({\frac{B}{A+2B\Phi}}\right)x^{a}x^{b}\right]\]
so that, with some effort, \[
\Gamma_{ab}^{k}={\frac{1}{2A}}H_{1}-\left({\frac{B}{2A(A+2B\Phi)}}\right)H_{2}\]
where 
\[
H_{1}  =  A'(x^{a}\delta_{bk}+x^{b}\delta_{ak}-x^{k}\delta_{ab}) 
+  B'x^{a}x^{b}x^{k}+2B\delta_{ab}x^{k}  \]
and 
\[
H_{2}  =  A'(2x^{a}x^{b}x^{k}-2\Phi x^{k}\delta_{ab})
 +  2\Phi B'x^{a}x^{b}x^{k}+4\Phi Bx^{k}\delta_{ab}. \]
Consequently, 
\[
g_{ab}  =  {\frac{\partial^{2}M}{\partial x^{a}\partial x^{b}}}-\Gamma_{ab}^{k}{\frac{\partial M}{\partial x^{k}}}\]
\[\equiv\delta_{ab}M'\left({\frac{A+A'\Phi}{A+2B\Phi}}\right)
 +  x^{a}x^{b}\left(M''-M'\left({\frac{A'+B'\Phi}{A+2B\Phi}}\right)\right). \]
Comparison with (\ref{A1}) now leads directly to 
\begin{equation}
A  =  M'\left({\frac{A+A'\Phi}{A+2B\Phi}}\right)=M'\left(\frac{(A\Phi)'}{A+2B\Phi}\right),\label{mass1} \end{equation}
\begin{equation}
B  =  M''-M'\left({\frac{A'+B'\Phi}{A+2B\Phi}}\right)\label{mass2} 
\end{equation}
which is a second order differential equation for the determination of $M(R)$. 
\\
\\
The first of the two equations above can be rearranged as 
\begin{equation}
B= -{\frac{A}{2\Phi}}+{\frac{M'}{2\Phi}}\left(\frac{(A\Phi)'}{A}\right) \label{mass3}
\end{equation}
or as 
\begin{equation}
\left(\frac{M'}{A+2B\Phi}\right)={\frac{A}{(A\Phi)'}}, \label{mass4}
\end{equation}
and these expressions can be used to eliminate $B$ in the second
equation as follows.
\\
\\
Use of (\ref{mass4}) in (\ref{mass2}) gives
\[
B = M'' - \frac{A}{(A\Phi)'}\left(A'+B'\Phi \right) \rightarrow \]
\[
\left(A \Phi \right)'B+\left(A \Phi \right) B'= M'' \left(A \Phi \right)'-A A' 
\rightarrow \]
\begin{equation}
\left(A  \Phi B\right)' = M'' \left(A \Phi \right)' -A A' . \label{mass5}
\end{equation}
But, from (\ref{mass3}),
\[
A  \Phi B =  - \frac{1}{2}A^2+\frac{1}{2}M' \left(A \Phi \right)' 
\]
so that (\ref{mass5}) becomes:
\[
\left[ - \frac{1}{2}A^2+ \frac{1}{2}M' \left(A \Phi \right)' \right]' =  M'' \left(A \Phi \right)' -A A' \rightarrow \]
\[
\frac{1}{2} M'' \left(A \Phi \right)'+\frac{1}{2}M'\left( A \Phi\right)''
= M'' \left(A \Phi \right)'  \rightarrow \]
\[
M'\left( A \Phi\right)''
= M'' \left(A \Phi \right)' \rightarrow \]
\[
\left( d_0 M'\right) =  \left( A \Phi\right)' \rightarrow \]
\begin{equation}
d_0 \left( M - M_0 \right)
= A \Phi ~~~{\rm where}~~ M_0 \equiv M(0). \label{mass6}
\end{equation}
But, by (\ref{eqn0}), since $R\underset{\mathrm{def}}{=}G(m)$ where $G(0)=0$ by definition, then $M_0 =0$ by definition. Consequently, using (\ref{mass6}) and (\ref{mass3}), we find for $A$ and $B$ respectively:
\[
A \equiv {\frac{d_{0} M}{\Phi}}, \]
\[
B \equiv -{\frac{A}{2\Phi}}+\left(\frac{M'}{2\Phi}\right)\left(\frac{d_0 M'}{A} \right)\]
 \[=-\left(\frac{d_0 M}{2\Phi^{2}}-\frac{M'M'}{2 M}\right)
\]
where we remember that, up to this point, we have been using the notation $M' \equiv d M/d\Phi$, where $\Phi \equiv R^2/2$.
We can now revert to the notation of the main text $\mathcal{M}'\equiv d M/dR$ etc, so that the foregoing results can be expressed as:
\[
A\equiv{\frac{2d_{0}M}{R^2}}, \]
\[B\equiv -\frac{1}{2R^2}\left(\frac{4 d_0 M}{ R^2}-\frac{M'M'}{ M}\right).
\]
\section{Conservative Form of Equations of Motion}\label{app.C} 
From the clock  constraint equation (\ref{(8A)}), we have
\begin{equation}
\mathbf{\dot{R}\cdot\dot{R}} = \frac{ c^2}{ 8 \pi \Sigma_F } A - \frac{B}{A} R^2\dot{R}^{2}. \label{B5}
\end{equation}
This suggests defining a potential function as: 
\begin{equation}
\mathcal{V} = E_{0}-\frac{1}{2}\left( \frac{ c^2}{8 \pi \Sigma_F } A-\frac{B}{A}R^2\dot{R}^{2}\right),\label{(10)}
\end{equation}
for some arbitrary constant $E_0$, and where $A$ and $B$ are defined at (\ref{4a}). Using the identity
\[
\frac{d\mathcal{V}}{dR} \equiv \frac{\partial \mathcal{V}}{\partial R}+\frac{\partial \mathcal{V}}{\partial\dot{R}}\frac{\ddot{R}}{\dot{R}},
\]
where we assume that the second term is always defined because the case of circular motions was \emph{explicitly} excluded in the derivation of (\ref{(5)}), then we easily obtain
\[
\nonumber \\
\frac{d\mathcal{V}}{dR}=- \frac{c^2}{16 \pi \Sigma_F} A'+\frac{R^2 \dot{R}^{2}}{2A}\left(B'-\frac{A'B}{A}\right)+\frac{B}{A}\left(R\dot{R}^{2}+R^{2}\ddot{R}\right). 
\]
The above expression
leads to 
\begin{equation}
2A\,\frac{d\mathcal{V}}{dR}\,\mathbf{\hat{R}}= \alpha \, \hat{\mathbf{R}}.
\label{B3}
\end{equation}
where
\[
\alpha \equiv \left(- \frac{c^2}{8 \pi \Sigma_F} AA'+B' R^2 \dot{R}^{2}-\frac{A'B}{A} R^2 \dot{R}^{2}+2B R\left(\dot{R}^2+R \ddot{R} \right)\right)
\]
From (\ref{B5}), we have 
\begin{equation}
\frac{ c^2}{ 8 \pi \Sigma_F } A = \frac{B}{A} R^2 \dot{R}^{2}+ \mathbf{\dot{R}\cdot\dot{R}} \label{B4}
\end{equation}
which, when substituted into (\ref{B3}), gives 
\[
2A\,\frac{d\mathcal{V}}{dR}\,\mathbf{\hat{R}}= \alpha \, \hat{\mathbf{R}}.
\]
where
\[
\alpha \equiv \left(B' R^2 \dot{R}^{2}+2B R \left(\dot{R}^2+R \ddot{R} \right)-A'\,\mathbf{\dot{R}\cdot\dot{R}}-2\frac{A'B}{A} R^2 \dot{R}^{2}\right)
\]
Finally, when used in (\ref{(9)}) this gives
\[
\mathbf{\ddot{R}} = - \frac{d\mathcal{V}}{dR}\,\mathbf{\hat{R}}
\]
for the result.
\section{Potential $\mathcal{V}$ purely in terms of $R$ and angular momentum} \label{C}
From the clock  constraint equation (\ref{ClockConstraint}), we have, directly,
\[
\dot{R}^2 + R^2 \dot{\theta}^2 = \frac{v_0^2}{8 \pi \Sigma_F } A  -\frac{B}{A} R^2\dot{R}^2
\]
from which we find
\[
\left(A+  B R^2\right) \dot{R}^2 + A  R^2\dot{\theta}^2 = \frac{v_0^2}{8 \pi \Sigma_F } A^2.
\]
A small amount of algebra then gives the clock  constraint equation as:
\[
\dot{R}^2 +R^2 \dot{\theta}^2 = \frac{v_0^2}{8 \pi \Sigma_F } \left(\frac{A^2}{A+ B R^2}\right) +h^2 \left(\frac{B}{A+ B R^2} \right)
\]
where, for conserved angular momentum, $h \equiv R^2 \dot{\theta}$. Consequently, the potential function of (\ref{(10)}) can be written as
\begin{equation}
\mathcal{V} =  -\frac{1}{2}\left[ \frac{v_0^2}{8 \pi \Sigma_F } \left(\frac{A^2}{A+ B R^2}\right) +h^2 \left(\frac{B}{A+  B R^2} \right)\right]  \label{C1}
\end{equation}
\\
\\
From (\ref{4a}), we have
\begin{equation}
A\equiv\frac{2\, d_0 M}{R^2},~~B\equiv-\frac{1}{2 R^2}\left(\frac{4\,d_0 M}{R^2}-\frac{M' M'}{ M}\right).\label{C3}
\end{equation}
so that
\[
A+ B R^2 = \frac{M' M'}{2 M} .
\]
A small amount of algebra now gives, for the expressions in (\ref{C1})
\[
\left(\frac{A^2}{A+ B R^2}\right) = \frac{8\, d_0^2 M^3}{ R^4  M' M'} \]
\[
\left(\frac{B}{A+ B R^2}\right) = -\frac{4\,d_0 M^2}{ R^4  M' M'}+ \frac{1}{R^2} \]
so that (\ref{C1}) becomes:
\[
\mathcal{V} =  -\frac{1}{2}\left[\frac{v_0^2}{8 \pi \Sigma_F } \left( \frac{8\, d_0^2 M^3}{ R^4  M' M'} \right) +h^2 \left( -\frac{4\,d_0 M^2}{ R^4  M' M'}+ \frac{1}{R^2} \right)  \right]
\]
and the corresponding form of the clock  constraint equation being:
\[
\dot{R}^2 +R^2 \dot{\theta}^2 = \left[ \frac{v_0^2}{8 \pi \Sigma_F } \left( \frac{8\, d_0^2 M^3}{ R^4  M' M'} \right) +h^2 \left( -\frac{4\,d_0 M^2}{ R^4  M' M'}+ \frac{1}{R^2} \right) \right] \]
\begin{equation}
= 4 \,d_0^2 v_0^2\left(\frac{  M}{4 \pi \Sigma_F } -\frac{ h^2 }{d_0 v_0^2} \right)\left( \frac{  \,M^2}{R^4 M' M'} \right)+\frac{h^2}{R^2}
\label{C0}
\end{equation}
In conventional terms this is equivalent to the energy equation.
%%%%%%%%%%%%%%%%%%%%%%%%%%%%%%%%%%%%%%%%%%%%%%%%%%

% Don't change these lines
%\bsp	% typesetting comment
%\label{lastpage}

\begin{thebibliography}{99}
	\bibitem[\protect\citeauthoryear{Alexander}{1984}]{Alexander1984}Alexander, H.G., \emph{The Leibniz-Clarke Correspondence,}
	Manchester University Press, 1984.
\bibitem[\protect\citeauthoryear{McKeon}{1941}]{McKeon1941}Aristotle, \emph{Categories,} in R. McKeon, \emph{The
	Basic Works of Aristotle,} Random House, New York, 1941.
\bibitem[\protect\citeauthoryear{Wicksteed \& Cornford}{1929}]{Wicksteed} Aristotle, \emph{Physics,} translated by P.H. Wicksteed \& F.M. Cornford, Leob Classical Library, Harvard University
Press, Cambridge, 1929
\bibitem[\protect\citeauthoryear{Baryshev et al}{1995}]{Baryshev1995} Baryshev, Yu V., Sylos Labini, F., Montuori, M., Pietronero, L. 1995 \textit{Vistas in Astronomy} 38, 419
\bibitem[\protect\citeauthoryear{Ayers}{1992}]{Berkeley1992} Berkeley, G., \emph{Of motion, or the principle and nature of motion and the cause of the communication of motions
(}Latin: \emph{De Motu)} in M.R.Ayers (ed) \emph{George Berkeley's Philosophical Works,} Everyman, London, 1992.
\bibitem[\protect\citeauthoryear{Broadhurst et al}{1990}]{Broadhurst}Broadhurst, T.J., Ellis, R.S., Koo, D.C., Szalay, A.S., 1990, Nat 343, 726
\bibitem[\protect\citeauthoryear{Charlier}{1908}]{Charlier1908}Charlier, C.V.L., 1908, Astronomi och Fysik 4,
1 (Stockholm)
\bibitem[\protect\citeauthoryear{Charlier}{1922}]{Charlier1922}Charlier, C.V.L., 1922, Ark. Mat. Astron. Physik
16, 1 (Stockholm)
\bibitem[\protect\citeauthoryear{Charlier}{1924}]{Charlier1924}Charlier, C.V.L., 1924, Proc. Astron. Soc. Pac. 37, 177
\bibitem[\protect\citeauthoryear{Da Costa et al}{1994}]{DaCosta} Da Costa, L.N., Geller, M.J., Pellegrini, P.S., Latham, D.W., Fairall, A.P., Marzke, R.O., Willmer, C.N.A., Huchra,
J.P., Calderon, J.H., Ramella, M., Kurtz, M.J., 1994, ApJ 424, L1
\bibitem[\protect\citeauthoryear{De Lapparent et al}{1988}]{DeLapparent1988} De Lapparent, V., Geller,M.J., Huchra, J.P., 1988, ApJ 332, 44
\bibitem[\protect\citeauthoryear{De Vaucouleurs}{1970}]{De Vaucouleurs1970} De Vaucouleurs, G., 1970, Sci 167, 1203
\bibitem[\protect\citeauthoryear{Gabrielli \& Sylos Labini}{2001}]{Gabrielli} Gabrielli, A., Sylos Labini, F., 2001, Europhys.
Lett. 54 (3), 286
\bibitem[\protect\citeauthoryear{Giovanelli and Haynes}{1986}]{Giovanelli1986} Giovanelli, R., Haynes, M.P., Chincarini, G.L., 1986, ApJ 300, 77
\bibitem[\protect\citeauthoryear{Hogg et al}{2005}]{Hogg} Hogg, D.W., Eistenstein, D.J., Blanton M.R., Bahcall N.A, Brinkmann, J., Gunn J.E., Schneider D.P. 2005 ApJ, 624, 54
\bibitem[\protect\citeauthoryear{Huchra et al}{1983}]{Huchra1983} Huchra, J., Davis, M., Latham, D.,Tonry, J., 1983, ApJS 52, 89
\bibitem[\protect\citeauthoryear{Joyce, Montuori \& Sylos Labini et al}{1999}]{Joyce} Joyce, M., Montuori, M., Sylos Labini, F., 1999,
Astrophys. J. 514, L5
\bibitem[\protect\citeauthoryear{Labini \& Gabrielli}{2000}]{Labini} Labini, F.S., Gabrielli, A., 2000, \emph{Scaling and fluctuations in galaxy distribution: two tests to probe large scale structures}, astro-ph0008047
\bibitem[\protect\citeauthoryear{Ariew \& Garber}{1989}]{Ariew} Leibniz, G.W., \emph{Philosophical Essays}, Edited
and translated by R. Ariew \& D.Garber, Hackett Publishing Co, Indianapolis,
1989
\bibitem[\protect\citeauthoryear{Lelli, McGaugh \& Schombert}{2015}]{McGaugh2015} Lelli, F., McGaugh, SS, Schombert, JM., \emph{The small scatter of the baryonic Tully-Fisher relation}, arxiv.org/abs/1512.04543
\bibitem[\protect\citeauthoryear{Lelli, McGaugh \& Schombert}{2016A}]{Lelli2016A} Lelli, F., McGaugh, SS, Schombert, JM., Ap. J.,  152, 6, 2016A
\bibitem[\protect\citeauthoryear{Lelli, McGaugh \& Schombert}{2016B}]{Lelli2016B} Lelli, F., McGaugh, SS, Schombert, JM., Ap. J.L.,  816, L14, 2016B
\bibitem[\protect\citeauthoryear{Mach}{1919}]{Mach1960}Mach, E., 1919, \emph{The Science of Mechanics} - \emph{a Critical and Historical Account of its Development} Open Court, La
Salle, 1960
\bibitem[\protect\citeauthoryear{Martinez et al}{1998}]{Martinez} Martinez, V.J., PonsBorderia, M.J., Moyeed, R.A., Graham, M.J. 1998 \textit{MNRAS} 298, 1212 
\bibitem[\protect\citeauthoryear{Milgrom}{1983a}]{Milgrom1983a}	Milgrom, M., 1983a, Ap. J. 270: 365. 
\bibitem[\protect\citeauthoryear{Milgrom}{1983b}]{Milgrom1983b} Milgrom, M., 1983b, Ap. J. 270: 371
\bibitem[\protect\citeauthoryear{Milgrom}{1983c}]{Milgrom1983c} Milgrom, M. 1983c. Ap. J. 270: 365-370
\bibitem[\protect\citeauthoryear{Milgrom}{1983d}]{Milgrom1983d} Milgrom, M. 1983d. Ap. J. 270: 371-383
\bibitem[\protect\citeauthoryear{Milgrom}{1983e}]{Milgrom1983e} Milgrom, M. 1983e. Ap. J. 270: 384-389

\bibitem[\protect\citeauthoryear{Newton}{1687}]{Newton1999}Newton, I., 1687, \emph{Principia: Mathematical Principles of Natural Philosophy,} translated by I.B. Cohen \& A. Whitman, University of California Press, 1999
\bibitem[\protect\citeauthoryear{Peebles}{1980}]{Peebles1980} Peebles, P.J.E., 1980, The Large Scale Structure of the Universe, Princeton University Press, Princeton, NJ.
\bibitem[\protect\citeauthoryear{Pietronero \& Sylos Labini}{2000}]{Pietronero} Pietronero, L., Sylos Labini, F., 2000, Physica
A, (280), 125\textsl{}
 \bibitem[\protect\citeauthoryear{Roscoe}{2022}]{Roscoe2022} Roscoe, D.F., http://arxiv.org/abs/2111.01700
\bibitem[\protect\citeauthoryear{Sachs}{1982}]{Sachs1} Sachs, M., \emph{General Relativity and Matter,}
Reidal, Dordrecht, 1982
\bibitem[\protect\citeauthoryear{Sachs}{1986}]{Sachs2} Sachs, M., \emph{Quantum Mechanics from General
	Relativity,} Reidal, Dordrecht 1986
\bibitem[\protect\citeauthoryear{Sanders}{2014}]{Sanders2014} Sanders, R. H., 2014, Canadian Journal of Physics. 93 (2): 126
\bibitem[\protect\citeauthoryear{Sciama}{1953}]{Sciama} Sciama, D.W., 1953, \emph{On the Origin of Inertia,}
Mon. Not. Roy. Astron. Soc, 113, 34
\bibitem[\protect\citeauthoryear{Scaramella et al}{1998}]{Scaramella} Scaramella, R., Guzzo, L., Zamorani, G., Zucca,
E., Balkowski, C., Blanchard, A., Cappi, A., Cayatte, V., Chincarini,
G., Collins, C., Fiorani, A., Maccagni, D., MacGillivray, H., Maurogordato,
S., Merighi, R., Mignoli, M., Proust, D., Ramella, M., Stirpe, G.M.,
Vettolani, G. 1998 \textit{A\&A} 334, 404 
\bibitem[\protect\citeauthoryear{Sylos Labini \& Montuori}{1998}]{SylosLabini} Sylos Labini, F., Montuori, M., 1998, Astron. \& Astrophys., 331, 809
\bibitem[\protect\citeauthoryear{Sylos Labini, Montuori \& Pietronero}{1998}]{SylosLabini1} Sylos Labini, F., Montuori, M., Pietronero, L.,
1998, Phys. Lett., 293, 62
\bibitem[\protect\citeauthoryear{Sylos Labini, Vasilyev \& Baryshev}{2006}]{SylosLabini2} Sylos Labini, F., Vasilyev, N.L., Baryshev,
Y.V., Archiv.Astro.ph/0610938 
\bibitem[\protect\citeauthoryear{Tekhanovich \& Baryshev}{2016}]{Tekhanovich} Tekhanovich D.I.I and Baryshev Yu.V., Archiv.Astro.ph/1610.05206
\bibitem[\protect\citeauthoryear{Vettolani et al}{1993}]{Vettolani} Vettolani, G., et al., 1993, in: Proc. of Schloss Rindberg Workshop: Studying the Universe With Clusters of Galaxies
\bibitem[\protect\citeauthoryear{Wu, Lahav \& Rees}{1999}]{Wu} Wu, K.K.S., Lahav, O., Rees, M.J., 1999, Nature
397, 225
\end{thebibliography}
\end{document}